\documentclass{ws-ijmpa}

\begin{document}

\markboth{P. K. Srivastava}
{Description of Strongly Interacting Matter in A Hybrid Model}

\catchline{}{}{}{}{}
%
%%%%%%%%%%%%%%%%%%%%%%%%%%%%%%%%%%%%%%%%%%%%%%%%%%%%%%%%%%%%%%%%%%%%

\title{Strongly Interacting Matter at Finite Chemical Potential : Hybrid Model Approach}

\author{P. K. SRIVASTAVA}

\address{Department of Physics, Banaras Hindu University, Address\\
Varanasi, 221005,
INDIA\\
prasu111@gmail.com}

\author{C. P. SINGH}

\address{Department of Physics, Banaras Hindu University, Address\\
Varanasi, 221005,
INDIA\\
cpsingh\_bhu@yahoo.co.in}

\maketitle

\begin{history}
\received{Day Month Year}
\revised{Day Month Year}
\end{history}

\begin{abstract}
Search for a proper and realistic equation of state (EOS) for strongly interacting matter used in the study of the QCD phase diagram still appears as a challenging problem. Recently, we constructed a hybrid model description for the quark gluon plasma (QGP) as well as hadron gas (HG) phases where we used an excluded volume model for HG and a thermodynamically consistent quasiparticle model for the QGP phase. The hybrid model suitably describes the recent lattice results of various thermodynamical as well as transport properties of the QCD matter at zero baryon chemical potential ($\mu_{B}$). In this paper, we extend our investigations further in obtaining the properties of QCD matter at finite value of $\mu_{B}$ and compare our results with the most recent results of lattice QCD calculation. 
%Finally we demonstrate the existence of two different limiting energy regimes and propose that the connection point of these two limiting regimes would foretell the existence of critical point (CP) of the deconfining phase transition from HG to QGP in our hybrid model.
\end{abstract}

\ccode{PACS numbers:12.38.Mh, 12.38.Gc, 25.75.Nq, 24.10.Pa}

\section{Introduction}	
\noindent
Quantum chromodynamics (QCD) predicts that at sufficiently high temperatures ($T$) and/or chemical potentials ($\mu_{B}$), strongly interacting matter goes through a phase transition from colour insulating hadron gas (HG) phase to colour conducting quark gluon plasma (QGP) phase ~\cite{a,b,b1}. This is known as the deconfinement phase transition. Polyakov loop which is related to the free energy of quark is known as the order parameter of this deconfining phase transition~\cite{satz}. The other important property of QCD is spontaneous chiral symmetry breaking. This feature of chiral symmetry gives rise to a second type of phase transition known as chiral symmetry restoration phase transition at high temperature and/or density~\cite{satz}. Chiral condensate is a suitable order parameter to determine this kind of phase transition occur in QCD matter. Heavy ion collisions provide a unique opportunity to study this QCD phase transition from HG to QGP~\cite{b1}. A remarkable outcome of these experiments is the observation of a hot, strongly interacting matter which behaves much like a nearly perfect relativistic fluid~\cite{c,m}. Our endeavour in such studies is the search of a suitable equation of state (EOS) for the description of both phases of strongly interacting matter. Significant success has been gained in lattice calculations using QCD thermodynamics to provide a valid EOS for QCD matter at zero baryon chemical potential ~\cite{c1,c11,c2,c3,c4,c5,c55,c6}. However, the lattice methods still lack a reliability to describe the properties of matter possessing a finite density of baryons. Therefore, finding an EOS for QCD matter valid at zero as well as non-zero chemical potential is still a challenging problem. Consequently, in the absence of a proper and realistic EOS for QCD matter, precise mapping of the entire QCD phase boundary and finding the location of a hypothesized QCD critical point (CP) appears as a tough challenge before the experimentalists and theorists~\cite{d,e,f}.

 Recently, we constructed a hybrid model for HG and QGP phases where we made use of a new excluded volume model for HG and a thermodynamically consistent quasiparticle model for the QGP phase~\cite{g,h}. We have compared various thermodynamical and transport properties at $\mu_{B}=0$ as obtained in our hybrid model description with the recent lattice results and found an excellent agreement between them~\cite{g,h}. We have also constructed a deconfining first order phase boundary between HG and QGP and found an end point where the boundary terminates~\cite{h,i}. Further, we have shown that at this end point, the nature of phase transition changes as $\Delta s/T^{3}=(s_{QGP}-s_{HG})/T^{3}$ and $\Delta c_{s}^{2}=(c_{s}^{2})_{QGP}-(c_{s}^{2})_{HG}$ become zero where $s$ is the entropy density and $c_{s}^{2}$ is the square of speed of sound~\cite{g}. We have also found a cusp like behaviour in the variation of $\eta/s$ (where $\eta$ is the shear viscosity) with respect to temperature at the end point obtained in our hybrid model~\cite{g} which was, therefore, suggested as a hint for the existence of QCD CP by some authors~\cite{j,k1} earlier. All these results suggest that this end point could be taken as a CP of deconfinement phase transition from HG to QGP. The physical mechanism involved in this calculation is intuitively analogous to the percolation model where a first order phase transition results due to 'jamming' of baryons which thus restricts the mobility of baryons~\cite{k,q}. However, in the percolation model we do not have any comparison to what we should get in the QGP picture. Here we use a similar picture and we explicitly and separately consider both the phases, i.e., HG as well as QGP and hence it gives a clear understanding how a first-order deconfining phase transition can be constructed in nature and finally we reach to an interesting finding that the baryonic size is crucially responsible for the existence of CP on the phase boundary in such a construction. At low baryon density, overlapping mesons fuse into each other and form a large bag or cluster, whereas at high baryon density, hard-core repulsion among baryons, restricts the mobility of baryons. Consequently we consider two distinct limiting regimes of HG, one beyond CP is a meson-dominant regime and the other is a baryon dominant region.

In this paper, we want to explore further the suitability of the hybrid model in predicting the properties of QCD matter having a finite value of $\mu_{B}$. As we have pointed out that a hot, strongly coupled and nearly perfect fluid was observed at RHIC where the viscosity is small enough, $\eta/s\simeq 1/4\pi$~\cite{eta,gale}, thus ideal fluid description can indeed be appropriate to explain the expansion history of the matter created during the heavy ion collisions~\cite{m,l} at least at RHIC and LHC. Ideal relativistic hydrodynamics is entropy conserving i.e., the ratio of the entropy and the particle number is expected to remain constant during expansion of the medium in heavy ion collision. The EOS of strongly interacting matter should indicate such hydrodynamic description of the medium. Very recently Wuppartel-Budapest (WB) collaboration gave the isentropic trajectories on the $T-\mu_{B}$ plane and calculated various thermodynamical as well as transport properties along these trajectories~\cite{n}. This lattice calculation involves Taylor's expansion method to circumvent the fermion sign problem which always exists at finite $\mu_{B}$. In the same paper~\cite{n}, authors have also shown the results for various thermodynamic quantities at $\mu_{L}=400$ MeV (where $\mu_{L}/3=\mu_{u}=\mu_{d}$ while strangeness chemical potential $\mu_{s}$ has been calculated by imposing the net strangeness neutrality condition i.e. $n_{s}=0$).  Therefore, it will be worthwhile to compare the results obtained by the collaboration with the corresponding results of our hybrid model and thus its suitability for the description of QCD matter can be tested.

We must emphasize the new results obtained by us in order to demonstrate the importance of the present paper. The utility of our hybrid model essentially at finite $\mu_{B}$ can be well tested with the recent lattice results~\cite{n} so that the model is regarded to provide the most valid description for the EOS of QGP in the entire $T-\mu_{B}$ region. We first draw the isentropic trajectories at some fixed $S/N$ (which is equivalent to fixed $S/N_{L}$ trajectories of Ref~\cite{n} since in our model we include the condition of net strangeness neutrality) on $T-\mu_{B}$ plane. We plot the isentropic trajectories having $S/N~=~300, ~100, ~45$ and $30$. Here $S/N~=300,~45,~30$ correspond to typical ratios at RHIC, SPS and AGS, respectively~\cite{o}. We also show the the location of CP on first order deconfining phase boundary as obtained earlier in our hybrid model~\cite{h,i} together with the chemical freezeout curve as obtained in our model of HG~\cite{p,r} on this $T-\mu_{B}$ plane to demonstrate its proximity to CP. We then calculate various thermodynamic as well as transport quantities like normalized pressure density ($p/T^{4}$), normalized energy density ($\epsilon/T^{4}$), normalized entropy density ($s/T^{3}$), trace anomaly ($\epsilon-3p/T^{4}$) and square of the speed of sound ($c_{s}^{2}$) and plot their variations with temperature along isentropic trajectory having $S/N~=~300$ which corresponds to the highest RHIC energy i.e., $\sqrt{s_{NN}}=200$ GeV. We compare our model results with the corresponding results obtained in recent lattice calculation~\cite{n}. We also calculate $p/T^{4}$, $\epsilon/T^{4}$ and $s/T^{3}$ for strongly interacting matter at $\mu_{B}=400$ MeV and show their variations with respect to temperature in the present hybrid model description which is again compared with lattice data. 

The rest of the paper is organised as follows : in Sec. 2 and 3, we will provide a brief description of our HG model and thermodynamically consistent quasiparticle model of QGP, respectively. Detailed description of these models can be found in our previous work~\cite{g,h,r}. Sec. 4 will demonstrate all the results obtained by us in our hybrid model and their comparison with the lattice results.

\section{Formulation of excluded volume model for HG}
\noindent
In the excluded volume models, repulsive interaction between two baryons has been included by giving the baryons a hard-core geometrical size and consequently reduces the hadronic degrees of freedom at large $T$ and/or $\mu_{B}$. Consequently, the hadronic pressure is reduced and one can get a deconfinement phase transition from HG to QGP by using Gibbs' construction. We  incorporate excluded-volume correction arising due to baryonic size only in our excluded volume model. We assume that mesons can overlap and fuse into one another and hence do not possess any hard-core repulsion. The grand canonical partition function for the baryons in HG, with full quantum statistics and after incorporating excluded volume correction can be explicitly written as~\cite{g,p,r}:
\begin{eqnarray}
ln Z_i^{ex} &=& \frac{g_i}{6 \pi^2 T}\int_{V_i^0}^{V-\sum_{j} N_j V_j^0} dV
\nonumber\\
&&\int_0^\infty \frac{k^4 dk}{\sqrt{k^2+m_i^2}} \frac{1}{[exp\left(\frac{E_i - \mu_i}{T}\right)+1]}
\end{eqnarray}
where $g_i$ is the degeneracy factor of ith species of baryons, $E_{i}$ is the energy of the particle ($E_{i}=\sqrt{k^2+m_i^2}$), $V_i^0$ is the eigenvolume of one baryon of ith species and $\sum_{j}N_jV_j^0$ is the total occupied volume by the baryons and $N_{j}$ represents total number of baryons of jth species.

Now we can write Eq.(1) as:

\begin{equation}
ln Z_i^{ex} = V(1-\sum_jn_j^{ex}V_j^0)I_{i}\lambda_{i},
\end{equation}
where $I_{i}$ represents the integral:
\begin{equation}
I_i=\frac{g_i}{6\pi^2 T}\int_0^\infty \frac{k^4 dk}{\sqrt{k^2+m_i^2}} \frac1{\left[exp(\frac{E_i}{T})+\lambda_i\right]},
\end{equation}
and $\lambda_i = exp(\frac{\mu_i}{T})$ is the fugacity of the particle, $n_j^{ex}$ is the number density of jth type of baryons after excluded volume correction and can be obtained from Eq.(2) as:
\begin{equation}
n_i^{ex} = \frac{\lambda_i}{V}\left(\frac{\partial{ln Z_i^{ex}}}{\partial{\lambda_i}}\right)_{T,V}
\end{equation}
This leads to a transcendental equation as
\begin{equation}
n_i^{ex} = (1-R)I_i\lambda_i-I_i\lambda_i^2\frac{\partial{R}}{\partial{\lambda_i}}+\lambda_i^2(1-R)I_i^{'}
\end{equation}
where $I_{i}^{'}$ is the partial derivative of $I_{i}$ with respect to $\lambda_{i}$ and $R=\sum_in_i^{ex}V_i^0$ is the fractional occupied volume. We can write R in an operator equation as follows~\cite{r,s,ss}:
\begin{equation}
R=R_{1}+\hat{\Omega} R
\end{equation}
where $R_{1}=\frac{R^0}{1+R^0}$ with $R^0 = \sum n_i^0V_i^0 + \sum I_i^{'}V_i^0\lambda_i^2$; $n_i^0$ is the density of pointlike baryons of ith species and the operator $\hat{\Omega}$ has the form :
\begin{equation}
\hat{\Omega} = -\frac{1}{1+R^0}\sum_i n_i^0V_i^0\lambda_i\frac{\partial}{\partial{\lambda_i}}
\end{equation}
Using Neumann iteration method and retaining the series upto $\hat{\Omega}^2$ term, we get
\begin{equation}
R=R_{1}+\hat{\Omega}R_{1} +\hat{\Omega}^{2}R_{1}
\end{equation}
\noindent
Eq.(8) can be solved numerically for R. Finally, we get the total pressure~\cite{r,s,ss} of the hadron gas:
\begin{equation}
\it{p}_{HG}^{ex} = T(1-R)\sum_iI_i\lambda_i + \sum_j\it{p}_j^{meson}
\end{equation}

In Eq. (9), the first term on the right hand side represents the pressure due to all types of baryons and the second term gives the total pressure from all mesons in HG. We use ideal gas description to calculate the mesonic pressure~\cite{r}. In this calculation, we have taken an equal volume $V^{0}=\frac{4 \pi r^3}{3}$ for each baryon with a hard-core radius $r=0.8$ fm. The reason to choose this hard-core radius has been discussed in ref.~\cite{g}. We have taken all baryons and mesons and their resonances having masses upto $2 GeV/c^{2}$ in our calculation of HG pressure. We have also used the condition of strangeness neutrality by putting $\sum_{i}S_{i}(n_{i}^{s}-\bar{n}_{i}^{s})=0$, where $S_{i}$ is the strangeness quantum number of the ith hadron, and $n_{i}^{s}(\bar{n}_{i}^{s})$ is the strange (anti-strange) hadron density of ith species, respectively. 

\section{Quasiparticle Model (QPM)}

The EOS for QGP, as used in this paper has been described in detail in reference~\cite{t,u}. The effective mass of the gluon in the quasiparticle model is ~\cite{g,h,t,u}:
\begin{equation}
m_{g}^{2}(T)=\frac{N_c}{6} g^{2}(T) T^{2} \left(1+\frac{N_{f}^{'}}{6}\right),
\end{equation}
\noindent
where $N_c$ represents the number of colours. We have taken $N_c=3$ in our calculation and:
\begin{equation}
N_{f}^{'}=N_{f}+\frac{3}{\pi^2}\sum_{f}\frac{\mu_f^2}{T^2}.
\end{equation}
\noindent
Here $N_f$ is the number of flavours of quarks and $\mu_f$ is the quark chemical potential belonging to the flavour f. We take the effective mass of the quarks as :
\begin{equation}
m_{q}^{2}=m_{q0}^{2}+\sqrt{2}m_{q0}m_{th}+m_{th}^{2},
\end{equation}
\noindent
were $m_{q0}$ is the rest mass of the quarks. $m_{th}$ represents the thermal mass of the quarks~\cite{g,h,t,u}:
\begin{equation}
m_{th}^{2}(T,\mu)=\frac{N_{c}^{2}-1}{8 N_{c}}\left[T^{2}+\frac{\mu_{q}^{2}}{\pi^2}\right]g^{2}(T),
\end{equation}
\noindent
Taking these values for the effective masses, energy density can be derived from the grand canonical  partition function in a thermodynamically consistent manner and is given as \cite{g,h}:
\begin{eqnarray}
\epsilon &=&\frac{T^4}{\pi^2}\sum_{l=1}^{\infty}\frac{1}{l^4}[\frac{d_g}{2}\epsilon_{g}(x_{g}l)+(-1)^{l-1}d_{q}cosh(\mu_{q}/T)\epsilon(x_{q}l)
\nonumber \\
&&+(-1)^{l-1}\frac{d_{s}}{2}\epsilon_{s}(x_{s}l)],
\end{eqnarray}
\noindent
with $\epsilon_{i}(x_{i}l)=(x_{i}l)^{3}K_{1}(x_{i}l)+3 (x_{i}l)^{2}K_{2}(x_{i}l)$, where $K_1$ and $K_2$ are the modified Bessel functions with $x_{i}=\frac{m_i}{T}$ and index i runs for gluons, up-down quarks q, and strange quark s. Here $d_i$ are the degeneracies associated with the internal degrees of freedom. Now, by using the thermodynamic relation $\epsilon=T\frac{\partial \it {p}}{\partial T}-\it {p}$, pressure of system at $\mu_{q}=0$ can be obtained as :
\begin{equation}
\frac{\it{p}(T,\mu_{q}=0)}{T}=\frac{\it{p}_0}{T_0}+\int_{T_0}^{T}dT \frac{\epsilon(T,\mu_{q}=0)}{T^2},
\end{equation}
\noindent
where $\it{p}_0$ is the pressure at a reference temperature $T_0$. We have used $\it{p}_{0}$=0 at $T_{0}$=100 MeV in our calculation. We get the pressure for a system at finite $\mu_{q}$ as :
\begin{equation}
\it{p}(T,\mu_{q})=\it{p}(T,0)+\int_{0}^{\mu_{q}}n_{q}d\mu_{q}.
\end{equation}
\noindent
where the expression for $n_{q}$ is:
\begin{equation}
n_{q}=\frac{d_{q}T^{3}}{\pi^2}\sum_{l=1}^{\infty}(-1)^{l-1}\frac{1}{l^3}sinh(\mu_{q}/T)I_{i}(x_{i}l)
\end {equation}
\noindent
with $I_{i}(x_{i}l)=(x_{i}l)^2 K_{2}(x_{i}l)$. Thus all the thermodynamical quantities can be obtained in a consistent way by using this model~\cite{bannur}. We have used the following expression for the $T$ and $\mu_{B}$ dependence of the coupling constant~\cite{g}
\begin{eqnarray}
&\alpha_{S}(T)=\frac{g^{2}(T)}{4 \pi}=\frac{6 \pi}{\left(33-2 N_{f}\right)\ln \left(\frac{T}{\Lambda_{T}}\sqrt{1+a\frac{\mu_{q}^{2}}{T^2}}\right)}\times
&\nonumber \\
& \displaystyle{\left(1-\frac{3\left(153-19 N_f \right)}{\left(33-2 N_f\right)^2}\frac{\ln \left(2 \ln \frac{T}{\Lambda_T}\sqrt{1+a\frac{\mu_{q}^{2}}{T^2}} \right)}{\ln \left(\frac{T}{\Lambda_{T}}\sqrt{1+a\frac{\mu_{q}^{2}}{T^2}}\right) }\right)}&,
\end{eqnarray}
where $\Lambda_{T}=115 MeV$ and $a=\frac{1}{\pi^{2}}$.

\begin{figure}[ht!]
\includegraphics[height=24em]{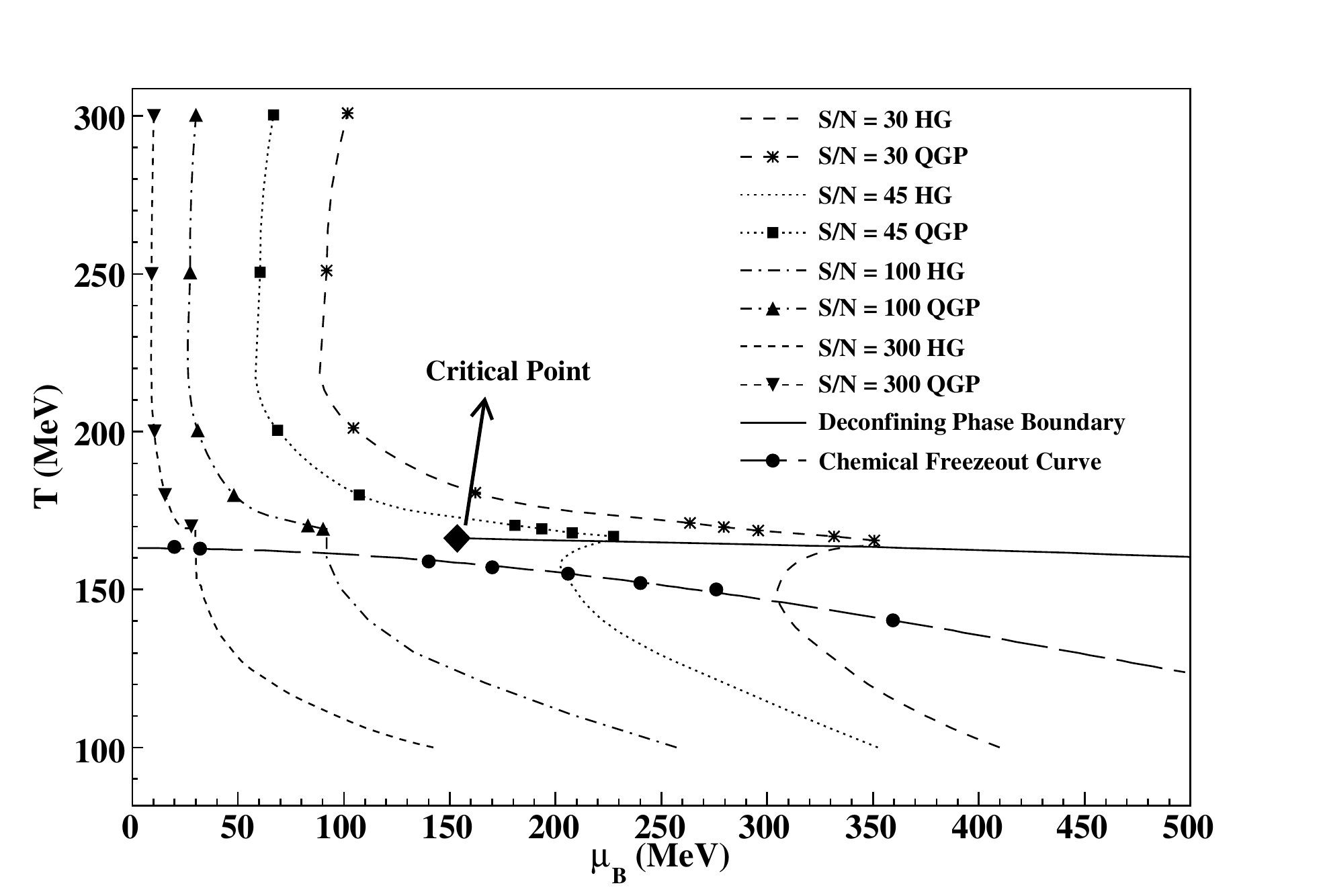}%use pdflatex for pdf
\caption[]{Isentropic trajectories in $T-\mu_{B}$ plane. First order deconfining phase boundary is shown by solid curve and solid diamond symbol represents the location of CP on this deconfining phase boundary between HG and QGP. Long dashed curve with solid circular symbols presents the chemical freeze-out curve obtained in our excluded volume model of HG.}
 \label{Fig:f1}
\end{figure}

\begin{figure}
\includegraphics[height=24em]{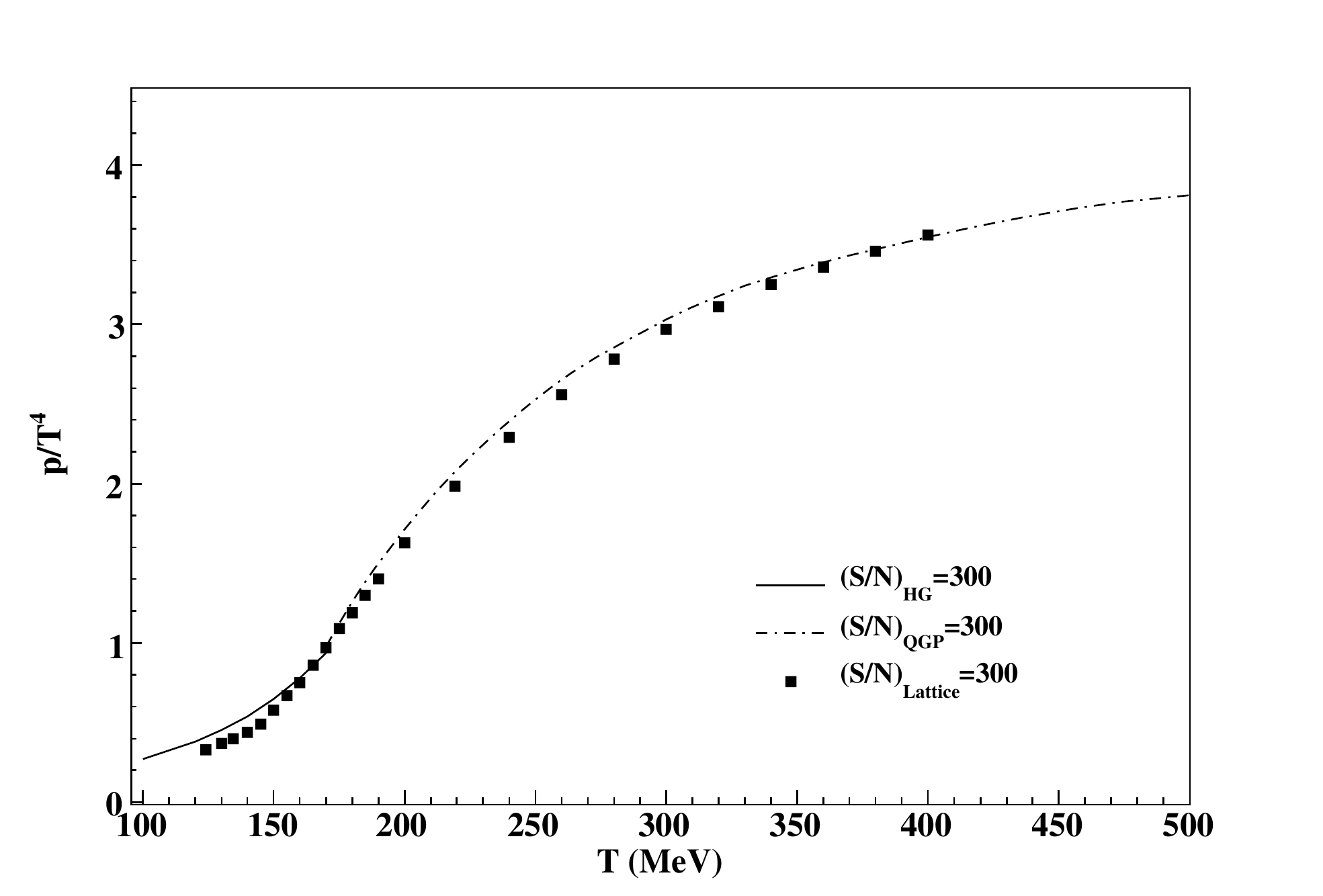}%use pdflatex for pdf
\caption[]{Variation of normalized pressure density with respect to temperature at constant $S/N~=~S/N_{L}=~300$ corresponding to RHIC energy (i.e. $\sqrt{s_{NN}}=200$ GeV)\cite{o}. Solid squares represent the recent lattice results\cite{n}. \label{f2}}
\end{figure}

\begin{figure}
\includegraphics[height=24em]{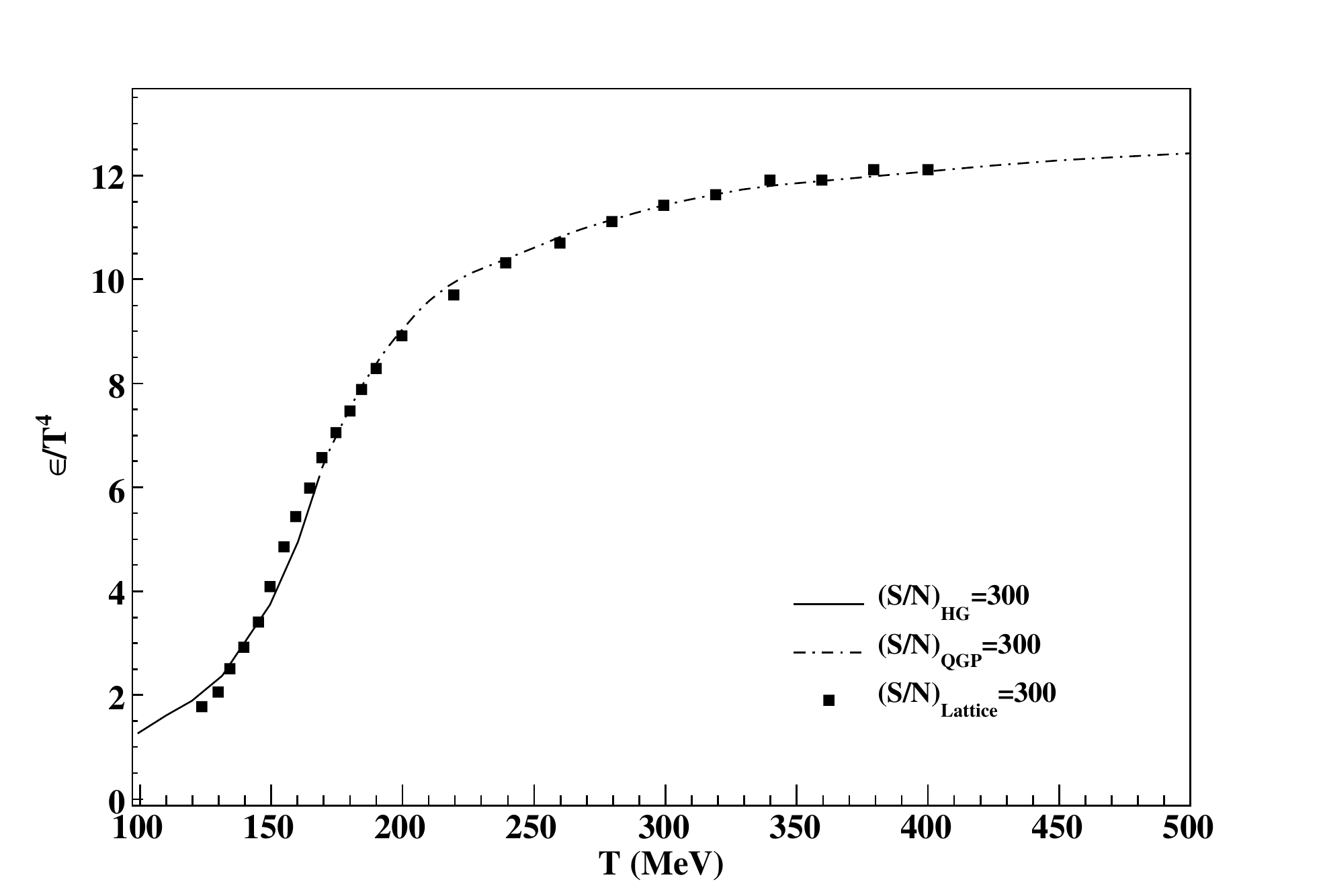}%use pdflatex for pdf
\caption[]{Variation of normalized energy density with respect to temperature at constant $S/N~=~S/N_{L}=~300$. Solid squares represent the recent lattice results\cite{n}. \label{f3}}
\end{figure}

\begin{figure}
\includegraphics[height=24em]{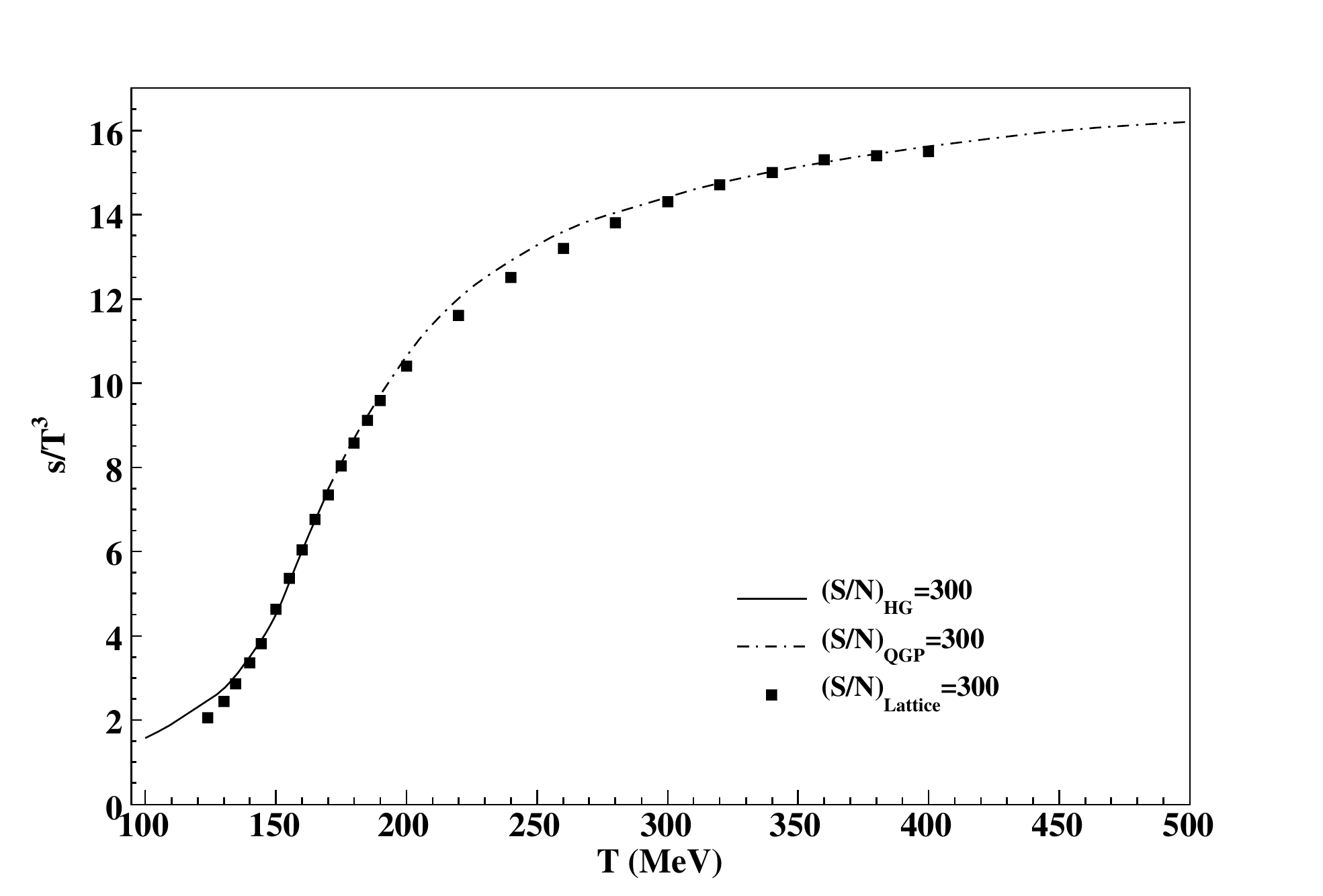}%use pdflatex for pdf
\caption[]{Variation of normalized entropy density with respect to temperature at constant $S/N~=~S/N_{L}=~300$. Solid squares represent the recent lattice results\cite{n}. \label{f4}}
\end{figure}

\begin{figure}
\includegraphics[height=24em]{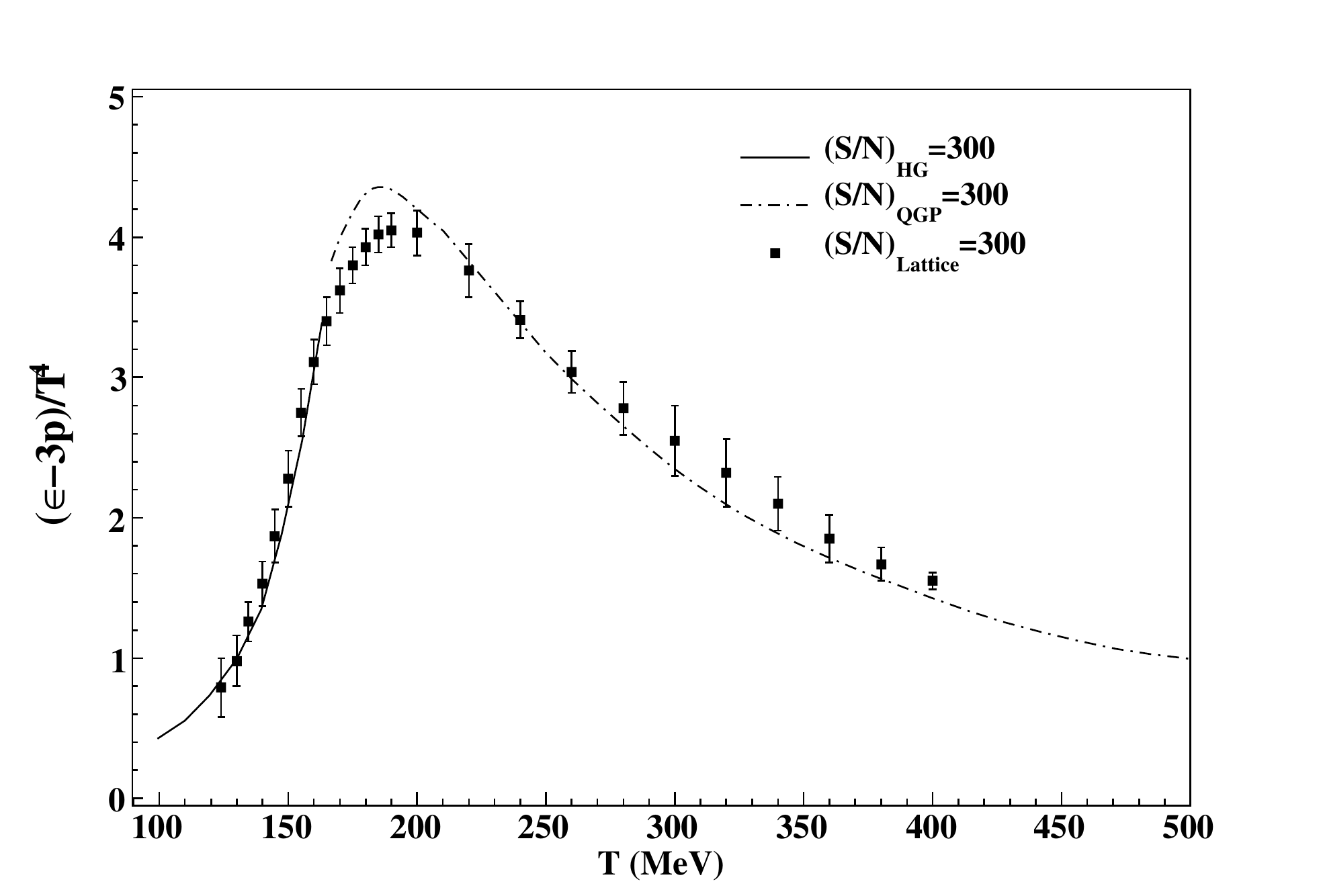}%use pdflatex for pdf
\caption[]{Variation of trace anomaly factor ($\epsilon - 3p/T^{4}$) with respect to temperature at constant $S/N~=~S/N_{L}=~300$. Solid squares represent the recent lattice results\cite{n}. \label{f5}}
\end{figure}

\begin{figure}
\includegraphics[height=24em]{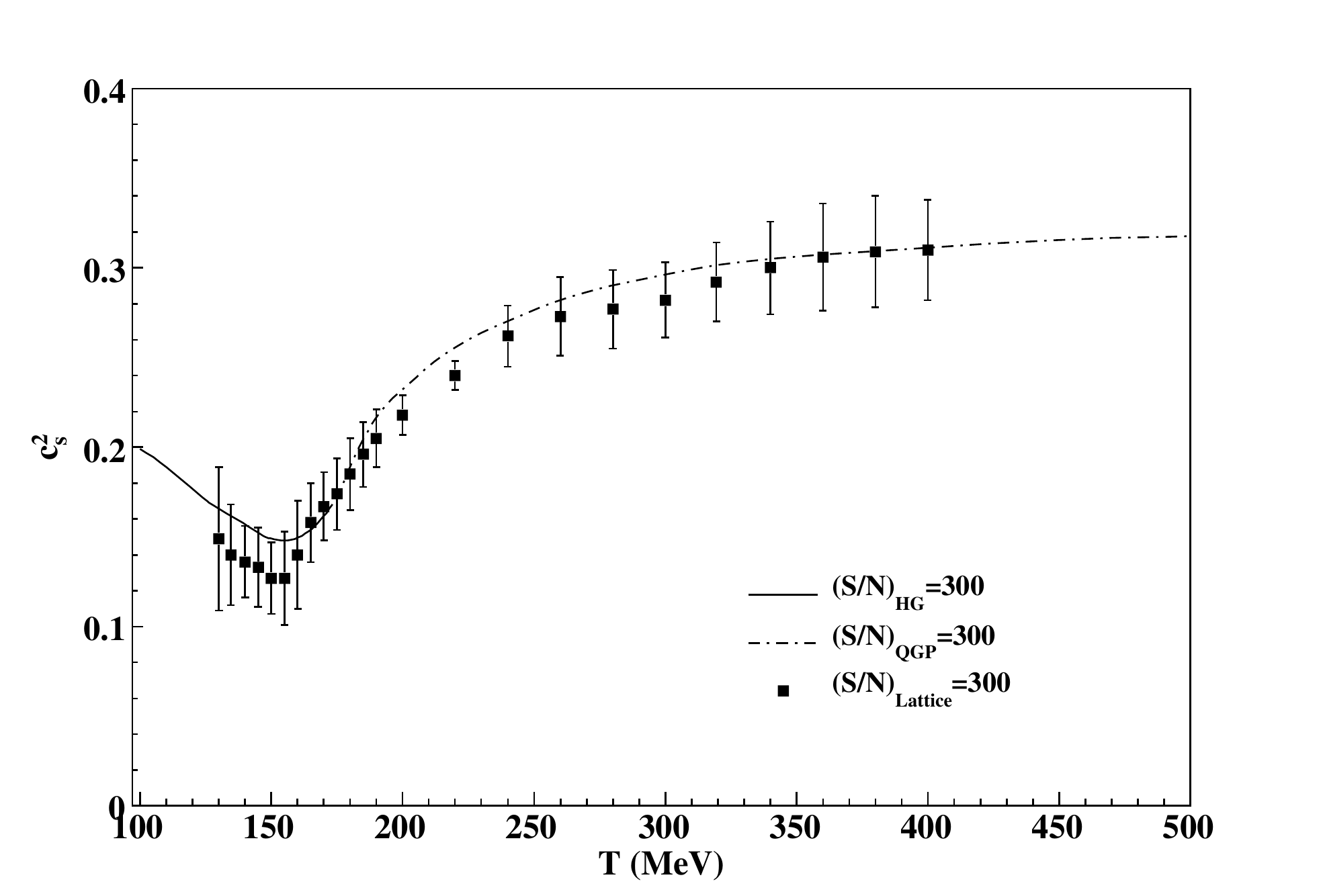}%use pdflatex for pdf
\caption[]{Variation of square of speed of sound ($c_{s}^{2}$) with respect to temperature at constant $S/N~=~S/N_{L}=~300$. Solid squares represent the recent lattice results\cite{n}. \label{f6}}
\end{figure}

\begin{figure}
\includegraphics[height=24em]{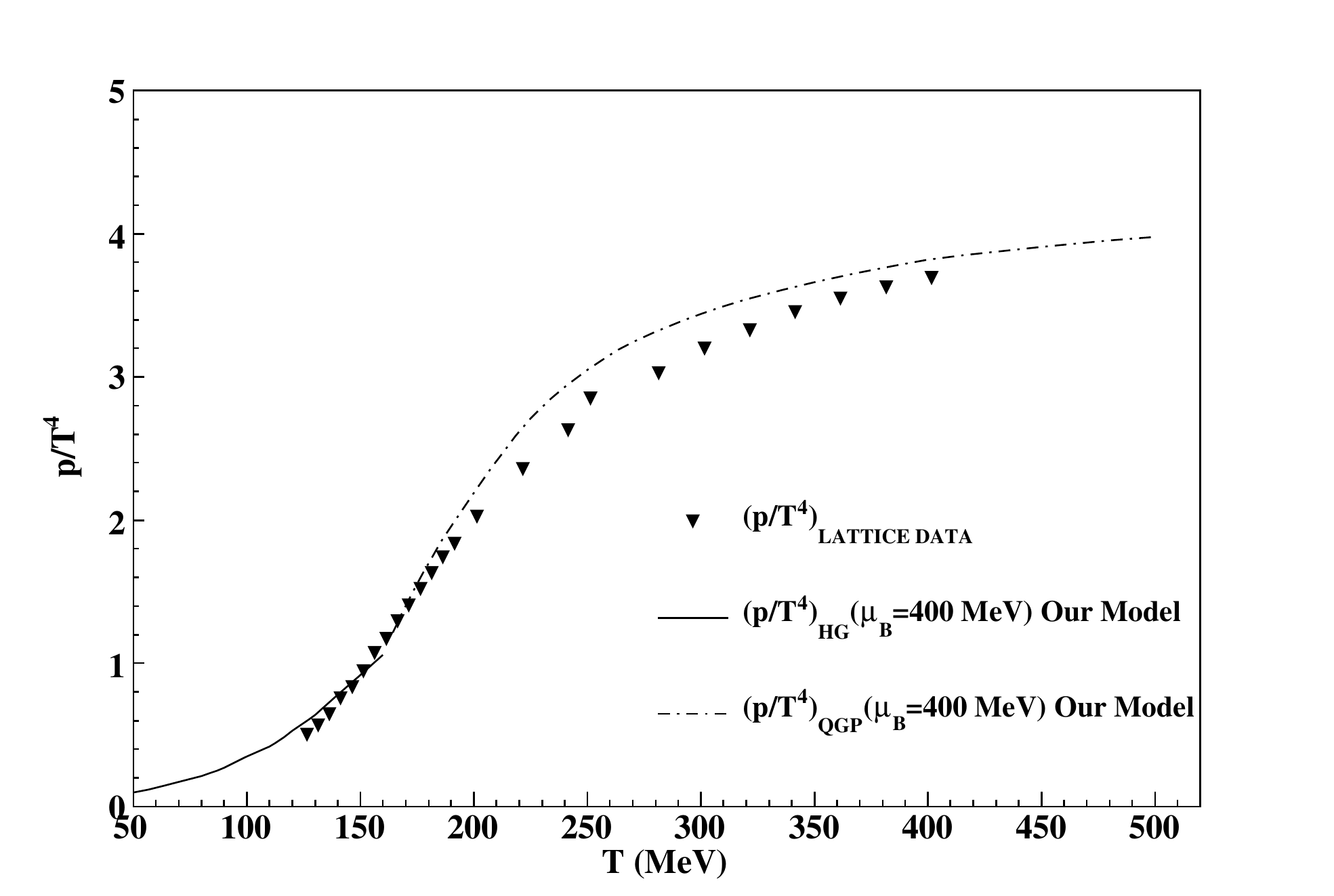}%use pdflatex for pdf
\caption[]{Variation of normalized pressure density with respect to temperature at $\mu_{B}=400$ MeV and comparison with lattice results\cite{n}. \label{f7}}
\end{figure}

\begin{figure}
\includegraphics[height=24em]{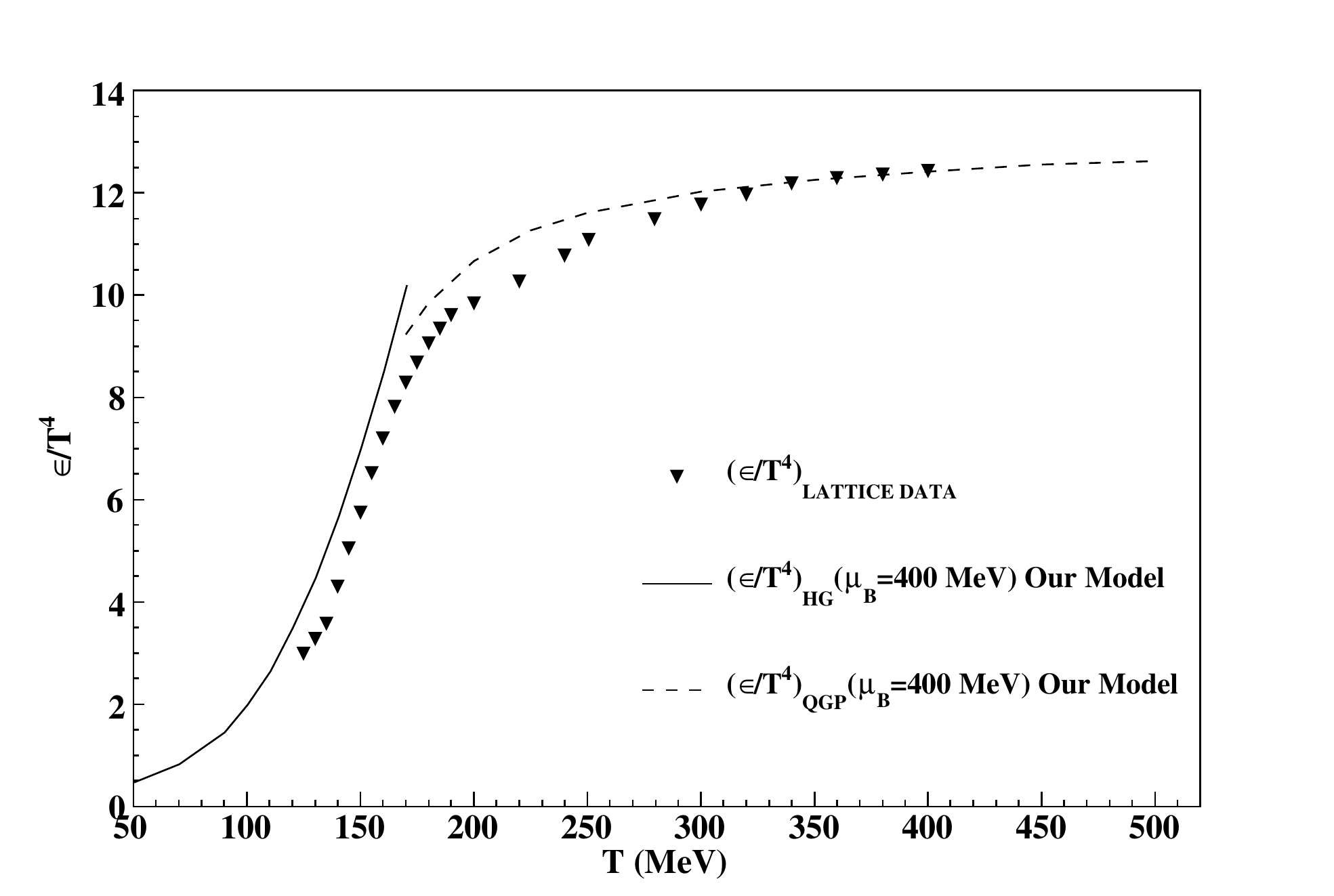}%use pdflatex for pdf
\caption[]{Variation of normalized energy density with respect to temperature at $\mu_{B}=400$ MeV and comparison with lattice results\cite{n}. \label{f8}}
\end{figure}

\begin{figure}
\includegraphics[height=24em]{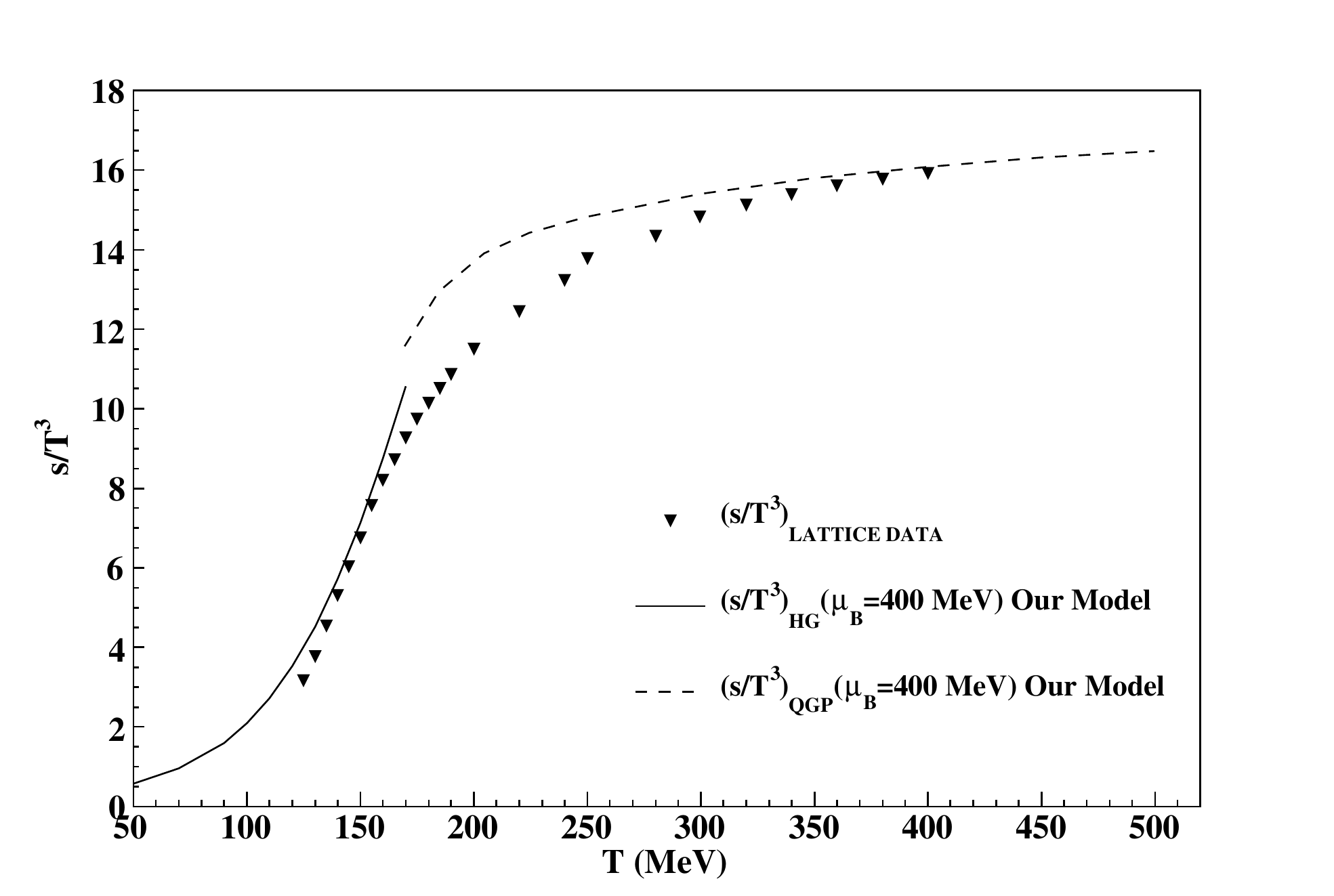}%use pdflatex for pdf
\caption[]{Variation of normalized entropy density with respect to temperature at $\mu_{B}=400$ MeV and comparison with lattice results\cite{n}. \label{f9}}
\end{figure}

\section{Results and Discussion}

 Fig.~\ref{Fig:f1} shows the trajectories for a given $S/N$ ($=S/N_{L}$ of Ref.~\cite{n}) in $T-\mu_{B}$ plane where $S$ is the total entropy, $N$ is the net particle number and $N_{L}$ is the net light particle number~\cite{n}. Here $N$ is equal to $N_{L}$ since we include net strangeness neutrality condition to calculate $\mu_{S}$. We also plot the first order deconfining phase boundary between HG and QGP phase as obtained in reference~\cite{h,i} . The end point of this phase boundary is shown by solid diamond symbol. We have shown earlier that the phase transition changes its nature from first order to higher order~\cite{g} at this end point and thus it represents a CP of deconfinement phase transition between HG and QGP phases. Furthermore, we draw the chemical freezeout curve as obtained in our excluded volume model for HG by fitting the experimental data of particle ratios~\cite{p,r}. Our CP corresponds to a value of $S/N=~60$ and it corresponds to a freezeout point at $T=158$ MeV and $\mu_{B}=~170$ MeV which further gives $\sqrt{s_{NN}}\approx~21$ GeV. From Fig.~\ref{Fig:f1}, it is clear that moving towards higher $S/N$ values means moving the trajectories from high $\mu_{B}$ region to low $\mu_{B}$ region. Here $S/N=~300$ thus corresponds to highest RHIC energy i.e., $\sqrt{s_{NN}}=~200$ GeV. 

Fig.~\ref{f2} demonstrates the variation of normalized pressure density ($p/T^{4}$) with respect to temperature for a trajectory at a constant $S/N$ ($=300$) as obtained from our hybrid model. This trajectory corresponds to $\sqrt{s_{NN}}=200$ GeV as mentioned earlier. We also compare our results with the corresponding results obtained in a recent lattice calculation~\cite{n} for the trajectory having the same value of $S/N$. We find that our hybrid model provides an excellent fit to the lattice data in the entire temperature range.

Similarly, Fig.~\ref{f3} and Fig.~\ref{f4} present the variations of normalized energy density ($\epsilon/T^{4}$) and normalized entropy density ($s/T^{3}$), respectively  with respect to temperature. We again compare our results with the corresponding results obtained in the lattice calculation~\cite{n} and find excellent agreement between them. We also observe that in our hybrid model result, there is no sign of any discontinuity at the joining point of HG and QGP phases in the variations of $\epsilon/T^{4}$ and $s/T^{3}$ with respect to temperatures. It is due to the passing of this constant $S/N$ trajectory from HG to QGP phase in crossover region as shown in Fig.~\ref{Fig:f1} in our hybrid model. Thus our model supports a smooth crossover from HG to QGP or vice-versa at highest RHIC energy and there is no hint for the occurrence of first order phase transition.

In Fig.~\ref{f5}, we show the variation of trace anomaly ($\epsilon-3p/T^{4}$) with respect to temperature in our hybrid model calculation. This quantity provides a measure of the interactions present in the medium. We also plot the lattice data obtained in ~\cite{n} for a comparison with our model results. We find that our hybrid model again appropriately describes the lattice data from very low to high temperatures. This study shows that the interaction present in the medium produced at highest RHIC energy is quite strong at a temperature corresponding to $170$ MeV. At higher temperature side, trace anomaly slowly decreases with the increase in temperature, however, it remains significant even upto $500$ MeV. Therefore, inclusion of interactions appears important in deriving the properties of the strongly interacting matter created at RHIC highest energy. Experimentalists have also found the signatures of strongly interacting QGP (sQGP) by measuring the elliptic flow at RHIC~\cite{c}.

Fig.~\ref{f6} reveals the variation of square of speed of sound ($c_{s}^{2}$) with respect to temperature at a constant $S/N$ ($=300$) obtained in our model. We find that it suitably matches with the $c_{s}^{2}$ given in lattice calculation~\cite{n}. We observe a shallow minimum in $c_{s}^{2}$ around $T\approx 160-170$ MeV which corresponds to a smooth crossover from HG to QGP phase or vice-versa as suggested in Refs.~\cite{v,w}. All the results at constant $S/N=~300$ suggest that the medium goes through a smooth crossover from HG to QGP at the highest RHIC energy i.e., $\sqrt{s_{NN}}=~200$ GeV around $T=~170$ MeV and the strength of the interaction is still quite significant.

Now, we want to calculate some thermodynamical quantities at a large value of $\mu_{B}$ ($=400$ MeV) and make comparisons with the corresponding results obtained in lattice calculation~\cite{n}. However, before making this exercise it is very important to note that for a large value of $\mu_{B}$, we need other lattice calculations separately in order to support the trends given here as the notorious sign problem creeping into lattice calculation and make it difficult at net baryon density (or baryon chemical potential). Moreover, the lattice EOS used here is primarily based on using Taylor expansion, and by construction it cannot reveal either a critical endpoint or first order phase transition. In Fig.~\ref{f7}, we have presented the variations of the normalized pressure density with respect to temperature at $\mu_{B}=400$ MeV. We have compared it with the lattice data points and found a good agreement with the results of our hybrid model in the low temperature region i.e., upto 225 MeV. However, at higher temperatures our hybrid model results are slightly higher than the lattice points.

Fig.~\ref{f8} demonstrates the variations of normalized energy density with respect to temperature at $\mu_{B}=400$ MeV. We find a good agreement at higher temperatures in contradiction to the normalized pressure density case (see Fig.~\ref{f7}). Also one noticeable thing is that we found a discontinuity at the joining point of HG and QGP phase in variation of $\epsilon/T^{4}$. However there is no sign of discontinuity present in lattice results. To make any clear statement one should wait for more refined lattice calculations.

Fig.~\ref{f9} shows the variation of $s/T^{3}$ with respect to $T$ at $\mu_{B}=400$ MeV. Lattice data points are also plotted to present a comparison with our model results. We find that at low as well as high temperatures our model results agree well with the lattice results. However, at intermediate temperature range i.e., from 160 MeV to 300 MeV, we find that our model results exceed over lattice results. We have also observed a discontinuity in our hybrid model results which clearly shows a first order phase transition at $T=160$ MeV for $\mu_{B}=400$ MeV in contradiction to lattice results which still shows a crossover type of behaviour at $\mu_{B}=400$ MeV.

In conclusion, we have constructed a proper and realsitic hybrid model for the description of strongly interacting QCD matter where we make use of an excluded volume model for HG and a thermodynamically consistent quasiparticle model for QGP phase. This model suitably works even in finite $\mu_{B}$ region and provides intuitive picture for crossover transition. We have also attempted to throw some light on the existence and the precise location of CP. We hope our results will be significant in future investigations regarding the EOS for QCD matter and the existence of CP on the QCD phase diagram. \\

\section*{Acknowledgments}
PKS is grateful to the University Grants Commission (UGC), New Delhi for providing a research grant.


\begin{thebibliography}{0}    %for 1 digit
\bibitem{a}
C. P. Singh, \emph{Phys. Rep.} {\bf 236} (1993) 147.

\bibitem{b}
H. Satz, \emph{Rep. Prog. Phys.} {\bf 63} (2000) 1511.

\bibitem{b1}
L. McLerran, \emph{Rev. Mod. Phys.} {\bf 58} (1986) 1021.

\bibitem{satz}
H. Satz, \emph{Int. J. Mod. Phys.} {\bf E 21}, (2012) 1230006.

\bibitem{c}
A. Adare et. al. [PHENIX Colla.], \emph{Phys. Rev. Lett.} {\bf 98} (2007) 172301.


\bibitem{m}
M. Gyulassy and L. McLerran, \emph{Nucl. Phys.} {\bf A 750} (2005) 30.


\bibitem{c1}
P. Huovinen and P. Petreczky, \emph{Nucl. Phys.} {\bf A 837} (2010) 26.

\bibitem{c11}
P. Huovinen and P. Petreczky, \emph{J. Phys. Conf. Ser.} {\bf 230} (2010) 012012.

\bibitem{c2}
C. Ratti et al., \emph{Nucl. Phys.} {\bf A 855} (2011) 253.

\bibitem{c3}
S. Plumari, W.M. Alberico, V. Greco, and C. Ratti, \emph{Phys. Rev.} {\bf D 84} (2011) 094004.

\bibitem{c4}
S. Borsanyi et al., \emph{JHEP} {\bf 11} (2010) 077.

\bibitem{c5}
S. Borsanyi et al., \emph{JHEP} {\bf 01} (2012) 138.

\bibitem{c55}
S. Borsanyi et al., \emph{J. Phys.} {\bf G 38} (2011) 124060.

\bibitem{c6}
M. Cheng et al., \emph{Phys. Rev.} {\bf D 81} (2010) 054504.

\bibitem{d}
M. A. Stephanov, \emph{Acta Physica Polonica} {\bf B 35} (2004) 2939. 

\bibitem{e}
M. A. Stephanov, \emph{Int. J. Mod. Phys.} {\bf A 20} (2005) 4387.

\bibitem{f} 
M. A. Stephanov, \emph{Phys. Rev. Lett.} {\bf 102} (2009) 032301.


\bibitem{g}
P. K. Srivastava and C. P. Singh, \emph{Phys. Rev.} {\bf D 85} (2012) 114016.

\bibitem{h}
P. K. Srivastava, S. K. Tiwari, and C. P. Singh, \emph{Phys. Rev.} {\bf D 82} (2010) 014023.


\bibitem{i}
P. K. Srivastava, S. K. Tiwari, and C. P. Singh, \emph{Nucl. Phys.} {\bf A 862-863CF} (2011) 424.


\bibitem{j}
C. Sasaki and K. Redlich, \emph{Nucl. Phys.} {\bf A 832} (2010) 62.

\bibitem{k1}
R. A. Lacey et. al., \emph{Phys. Rev. Lett.} {\bf 98} (2007) 092301.


\bibitem{k}
P. Castorina, K. Redlich and H. Satz, \emph{Eur. Phys. J.} {\bf C 59} (2009) 67. 

\bibitem{q}
A. Andronic et. al, \emph{Nucl. Phys.} {\bf A 837} (2010) 65.

\bibitem{eta}
P. Jacobs and X.-N. Wang, \emph{Prog. Part. Nucl. Phys.} {\bf 54} (2005) 443.

\bibitem{gale}
C. Gale, S. Jeon, B. Schenke, \emph{Int. J. Mod. Phys.} {\bf A 28} (2013) 1340011.

\bibitem{l}
U. W. Heing and P. F. Kolb, \emph{Nucl. Phys.} {\bf A 702} (2002) 269.

\bibitem{n}
Sz. Borsanyi et. al., \emph{JHEP} {\bf 08} (2012) 053.

\bibitem{o}
M. Bluhm, B. Kampfer, R. Schulze, D. Seipt, and U. Heinz,  \emph{Phys. Rev.} {\bf C 76} (2007) 034901.

\bibitem{p}
C. P. Singh, P. K. Srivastava and S. K. Tiwari, \emph{Phys. Rev.} {\bf D 80} (2009) 114508; \emph{Phys. Rev.} {\bf D 83} (2011) 039904(E).

\bibitem{r}
S. K. Tiwari, P. K. Srivastava, and C. P. Singh, \emph{Phys. Rev.} {\bf C 85} (2012) 014908.

\bibitem{s}
M. Mishra and C. P. Singh, \emph{Phys. Rev.} {\bf C 76} (2007) 024908.

\bibitem{ss}
M. Mishra and C. P. Singh, \emph{Phys. Rev.} {\bf C 78} (2008) 024910.

\bibitem{t}
V. M. Bannur, \emph{Phys. Lett.} {\bf B 647} (2007) 271.

\bibitem{u}
V. M. Bannur, \emph{Eur. Phys. J.} {\bf C 50} (2007) 629.

\bibitem{bannur}
V. M. Bannur, \emph{Int. J. Mod. Phys.} {\bf A 28} (2013) 1350006.

\bibitem{v}
M. Chojnacki and W. Florkowski, \emph{Acta Phys. Pol.} {\bf B 38} (2007) 3249 [arXiv:nucl-th/0702030].

\bibitem{w}
P. Castorina, J. Cleymans, D. E. Miller, H. Satz, (2009) [arXiv:0906.2289].

% Please avoid comments such as "For a review'', "For some examples",
% "and references therein" or move them in the text. In general,
% please leave only references in the bibliography and move all
% accessory text in footnotes.

% Also, please have only one work for each \bibitem.


\end{thebibliography}
\end{document}